\documentclass[letterpaper, 10 pt, conference]{ieeeconf}  

\usepackage{graphicx} 
\usepackage{xcolor}
\usepackage{colortbl}
\definecolor{gray}{gray}{0.9}

\usepackage{verbatim}
\usepackage{pgfplots}
\usepackage{colortbl}
\usepackage{pgfplotstable}
\usepgfplotslibrary{statistics}
\usepackage{booktabs}
\usepackage{subcaption}
\usepackage{hyperref}

\usepackage{lipsum}
\newcommand\blfootnote[1]{%
  \begingroup
  \renewcommand\thefootnote{}\footnote{#1}%
  \addtocounter{footnote}{-1}%
  \endgroup
}

\usepackage{xcolor}
\usepackage{colortbl}
\definecolor{gray}{gray}{0.9}
\definecolor{orange}{rgb}{1,0.5,0}
\definecolor{mdred}{rgb}{0.7,0,0}
\definecolor{mdgreen}{rgb}{0.05,0.6,0.05}
\definecolor{mdblue}{rgb}{0,0,0.7}
\definecolor{dkblue}{rgb}{0,0,0.5}
\definecolor{dkgray}{rgb}{0.3,0.3,0.3}
\definecolor{slate}{rgb}{0.25,0.25,0.4}
\definecolor{ltgray}{rgb}{0.7,0.7,0.7}
\definecolor{purple}{rgb}{0.7,0,1.0}
\definecolor{lavender}{rgb}{0.65,0.55,1.0}

\title{\LARGE \bf 
Navigating to Success in Multi-Modal Human-Robot Collaboration: Analysis and Corpus Release}

\author{
    {\bf Stephanie M. Lukin\textsuperscript{\rm 1}, Kimberly A. Pollard\textsuperscript{\rm 1}, Claire Bonial\textsuperscript{\rm 1}, Taylor Hudson\textsuperscript{\rm 2},}
    \\{\bf Ron Artstein\textsuperscript{\rm 3}, Clare Voss\textsuperscript{\rm 1}, and David Traum\textsuperscript{\rm 3}}\\
    \textsuperscript{\rm 1}DEVCOM Army Research Laboratory, Adelphi, MD 20783\\
    \textsuperscript{\rm 2}Oak Ridge Associated Universities, Oak Ridge, TN 37831\\
    \textsuperscript{\rm 3}USC Institute for Creative Technologies, Playa Vista, CA 90094\\
    \texttt{stephanie.m.lukin.civ@army.mil}
    }

\begin{document}

\maketitle
\thispagestyle{empty}
\pagestyle{empty}

\begin{abstract}

Human-guided robotic exploration is a useful approach to gathering information at remote locations, especially those that might be too risky, inhospitable, or inaccessible for humans. Maintaining common ground between the remotely-located partners is a challenge, one that can be facilitated by multi-modal communication. 
In this paper, we explore how participants utilized multiple modalities to investigate a remote location with the help of a robotic partner. Participants issued spoken natural language instructions and received from the robot: text-based feedback, continuous 2D LIDAR mapping, and upon-request static photographs. We noticed that different strategies were adopted in terms of use of the modalities, and hypothesize that these differences may be correlated with success at several exploration sub-tasks.
We found that requesting photos may have improved the identification and counting of some key entities (doorways in particular) and that this strategy did not hinder the amount of overall area exploration.
Future work with larger samples may reveal the effects of more nuanced photo and dialogue strategies, which can inform the training of robotic agents. Additionally, we announce the release of our unique multi-modal corpus of human-robot communication in an exploration context: {\it SCOUT}, the Situated Corpus on Understanding Transactions.\blfootnote{This is the author’s version of the work. It is posted here for your personal use. Not for redistribution. The definitive Version of Record was published in the Proceedings of the 2023 IEEE Robot and Human Interactive Communication Conference (RO-MAN ’23), August 28-31, 2023, Busan, Korea.}

\end{abstract}

\section{Introduction}

Human-guided robotic exploration is a useful approach to gathering information at remote locations, especially those that might be too risky, inhospitable, or inaccessible for humans. Maintaining common ground between the remotely-located partners is a challenge; one way to address it is
through multi-modal communication, where each modality contributes to and enables common ground. For instance, positional information may be conveyed through a dynamically updating 2D map, which can be enhanced with photos that show a snapshot of the robot's perspective. Language may be employed for issuing succinct or complex navigation instructions, influenced by the shared common ground observed on the map and through the photos. Within this rich context, we explore the characteristics and strategies humans used in a multi-modal communication exercise and how these relate to 
performance in area exploration and item-counting tasks.

\begin{figure}[h!]
    \centering
    \includegraphics[width=0.45\textwidth]{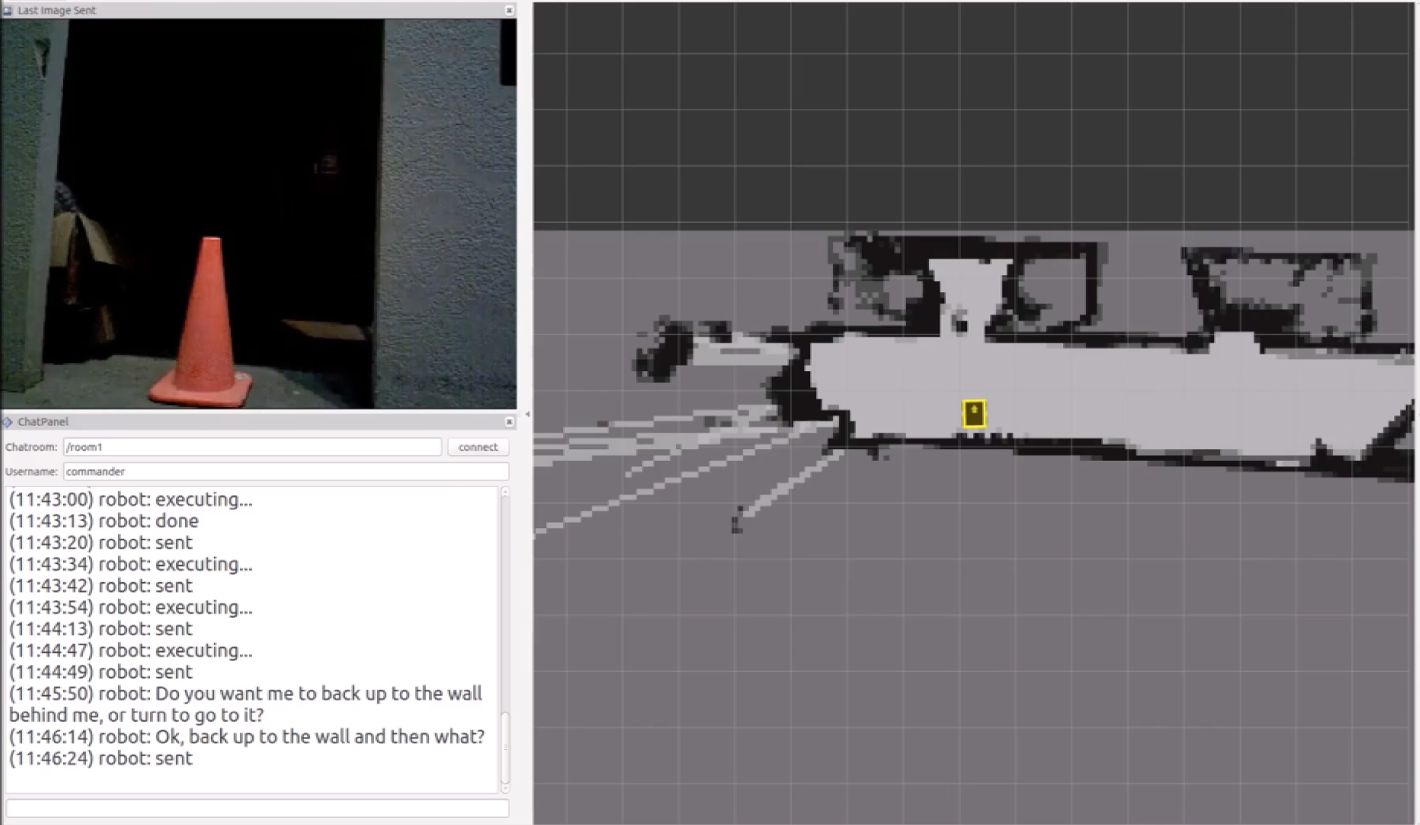}
    \caption{Participant's monitor display, showing dialogue exchange with the robot, LIDAR, and the last photo requested.}
    \label{exp-view}
\end{figure}

Our testbed for this exercise mimics a situation in which a human must work with their remotely-located robot teammate to explore an unknown area under low-bandwidth conditions---a situation representative of various inhospitable remote environments. This constraint prevents live-video streaming of the robot's perspective to the human or remote teleoperation of the robot. 
During the task, the human discovers, along with the robot, what the environment looks like by using language (as manifested in dialogue between the human and robot), photos (an on-demand window into the robot's perspective), and LIDAR (LIght Detection And Ranging, which progressively builds a spatial model displayed as a 2D floor plan to aid in robot navigation). The modalities are shown in Fig.~\ref{exp-view}. We are releasing the multi-modal data collected within this testbed in a corpus called {\it SCOUT}, the Situated Corpus on Understanding Transactions\footnote{We plan to make {\it SCOUT} available at \url{github.com/USArmyResearchLab} or by contacting the first author.}.

\begin{figure}[th!]
     \centering
  
     \begin{subfigure}{0.20\textwidth}
         \centering
         \includegraphics[width=\textwidth]{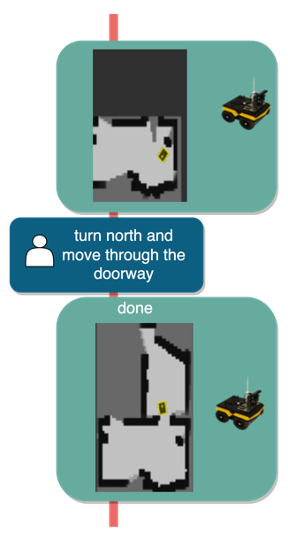}
         \vspace{-0.2in}
         \caption{Instruction from LIDAR}
         \label{dialogueA}
         
     \end{subfigure}
        \begin{subfigure}{0.25\textwidth}
         \centering
         \includegraphics[width=\textwidth]{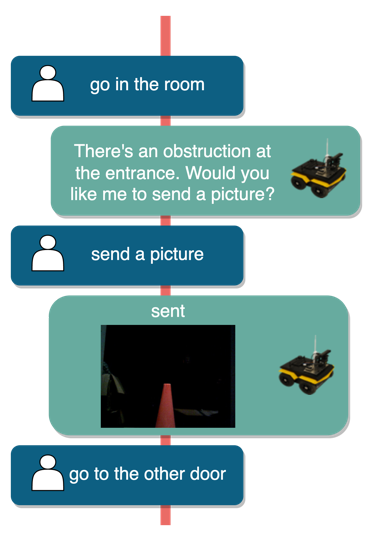}
         \vspace{-0.2in}
         \caption{Photo for replanning}
         \label{dialogueB}
     \end{subfigure}
     \caption{Multi-modal dialogue excerpts}
     \label{dialogue}
     \vspace{-0.2in}
  \label{figure}
\end{figure}

Some aspects of multi-modal communication in this environment can be seen in Fig.~\ref{dialogue}. Fig.~\ref{dialogueA} shows a LIDAR map at the moment in time before and after a human instructed a robot to ``turn north and move through the doorway.'' The cardinal direction is referring to the open doorway above the robot icon outlined in yellow, which is subsequently explored, thereby adding a new scanned area to the map post-execution.
A back and forth exchange is shown in Fig.~\ref{dialogueB} where a robot informs the human of an obstacle, and proposes that a photo will help establish visual common ground.

Within this exercise (described in Section~\ref{sec:exercise}) we define measures of success and hypothesize that these measures may be tied to participants' strategic use of one or more of the three modalities (language, LIDAR, photos). Success, as defined in Section~\ref{sec:success}, is scored by the completion of the item-counting tasks that the human was asked to perform based on the remote environment, and requires 
dialogue and photos to complete. Total area explored can also be considered a measure of success, as it can be critical in many remote exploration contexts. We examined sessions from ten participants and identified characteristics relating to how they took photos, how much they moved around the space, and how language was used to progress within the exercise (Section~\ref{sec:summary}). 

We test the following hypotheses on relationships between task success and the three modalities which enable dialogue, visual common ground, and exploration:

\begin{itemize}
    \item More exploration of the area should be correlated with better performance of item-counting tasks (H$_1$);
    \item Taking more photos should be correlated with better performance of item-counting tasks (H$_2$), as should more comprehensive photo requesting strategies (H$_3$), however there may be a trade-off between taking photos and exploring the area (H$_4$);
    \item Use of dialogue, as measured by the number of new intents issued throughout the exercise, should correlate with greater accomplishment in terms of better performance on item-counting tasks (H$_5$) and exploring the area (H$_6$);
    \item Use of dialogue may be related to the number of photos taken (H$_7$).
\end{itemize}

\noindent We review our findings in Section~\ref{sec:discussion} and situate them within the context of other human-robot exercises in Section~\ref{sec:related}. We conclude with ways in which Wizard-of-Oz (WoZ) 
and autonomous human-robot communication may be enhanced in our future work.

\begin{table*}[th!]
\centering
\caption{Dialogue excerpts across conversational floors \label{tab:tu}}
\centering
\begin{tabular}{|lp{0.2in}|p{1.4in}|p{1.4in}||p{1.4in}|p{1.0in}|}
\hline 
&&\multicolumn{2}{|c||}{{\it Left Floor}} & \multicolumn{2}{|c|}{{\it Right Floor}}\\ \hline 
TU\# & ID\# & {\bf Participant} & {\bf DM-Wizard$\rightarrow$Participant} & {\bf DM-Wizard$\rightarrow$RN-Wizard} & {\bf RN-Wizard} \\ \hline
1 & 1 & move to the right side of the orange cone & & & \\
\rowcolor{gray}1 & 2 & & & Move to the right side of the orange cone. & \\
1 & 3 & & executing... & & \\
\rowcolor{gray} 1 & 4 & & & & I don't think I'll fit \\
1 & 5 & & I won't fit there, but I can move further to the right, ok? & & \\ 
\rowcolor{gray} 1 & 6 & okay do that & & & \\ 
1 & 7 & & & move as far towards the right side as you can fit. &  \\  
\rowcolor{gray} 1 & 8 & & executing... & &  \\
1 & 9 & & & & done \\ 
\rowcolor{gray} 1 & 10 & & done & &  \\ \hline
2 & 11 & take a picture & & & \\  
\rowcolor{gray} 2 & 12 & & & photo & \\  
2 & 13 & & & & photo sent \\ 
\rowcolor{gray}2 & 14 & & sent & & \\ \hline
\end{tabular}
\end{table*}

\section{Human-Robot Collaboration Exercise}
\label{sec:exercise}

Our human-robot collaboration exercise was an exploration-and-search task in which human participants were given three resources: a robot, a map, and photographs \cite{marge2016applying}. The robot, a Clearpath Jackal, was located in a remote environment, and it  could understand spoken, natural language instructions from the participant (with the assistance of a WoZ confederate, see below). The verbal instructions were issued to the robot through a push-to-talk interface, and the robot responded through text messages. Communication occurred over wifi through a chain of Ubiquiti Bullet access points from the computer running the experimental software to the robot. The map was a 2D, birds-eye view LIDAR map of the environment using a Hokuyo LIDAR laser scanner, which dynamically updated as the robot moved through the space. Obstacles such as walls and large obstructions rendered in black, while open space appeared as light gray and unexplored space appeared as dark gray (see Figs.~\ref{exp-view} and \ref{dialogueA}). Finally, the photos were taken from the robot's front-facing RGB camera (Asus Xtion Pro Live), showing the view directly in front of the robot. Photos could be requested at any time. For the duration of the exercise, the participant, who assumed the role of `Commander,' sat at a workstation in an office room, with a microphone, keyboard, and a monitor display (an example is shown in Fig.~\ref{exp-view}).

Participants first completed a training trial to familiarize themselves with instructing the robot and the multiple modalities. They then completed two main trials in a house-like space with unfinished walls, floors, and sparse household items. Each trial started at a different location in the house, and the LIDAR map reset between trials. The participant was not informed that it was the same physical building. The trials lasted for 20 minutes or until the participant wanted to stop early. Participants were given a paper worksheet to complete with the help of their robot teammate: they had to count the number of doorways in the environment, which could be identified from photos or from LIDAR, and count objects of interest (shovels or shoes) and support having found the objects with photo evidence. Unlike doorways, these items could only be identified from photos. 

The exercise was carried out using a Wizard-of-Oz (WoZ) paradigm, in which key components of the robot were controlled by experimenters without revealing this to the participants. One Wizard handled the robot's dialogue understanding and processing capabilities as the Dialogue Manager (DM) Wizard, and the other handled the robot's movement as the Robot Navigator (RN) Wizard. The DM-Wizard listened to the participant's speech and typed responses and resolved miscommunication in what is called the {\it left conversational floor} \cite{traum2018dialogue}. Once the DM-Wizard deemed a participant's instruction as well-formed, the DM-Wizard translated the instruction into a more rigid natural language instruction and passed it to the RN-Wizard to execute. This took place in the {\it right conversational floor}, where only the DM-Wizard and RN-Wizard communicated. The RN-Wizard informed the DM-Wizard through spoken language to indicate completion or complication in executing an instruction; the DM-Wizard in turn informed the participant of this completion or complication via text. 
A dialogue excerpt is presented in Table~\ref{tab:tu}, which is annotated with the transaction unit (TU) schema defined in Traum et al. \cite{traum2018dialogue}. Utterances across the conversational floors are clustered into units that sequentially work towards fulfilling the original speaker's intent. A TU encompasses multiple dialogue turns, including requests for clarification and subsequent repairs, as seen in TU\#1 which begins with the participant's instruction to ``move to the right side of the orange cone'' (ID\#1) and ends when this action is completed and acknowledged (ID\#10). TU\#2 to follow contains the intent and fulfillment of taking a picture. 

\begin{figure*}[t]
    \centering
    \begin{subfigure}[t]{0.34\textwidth}
        \includegraphics[width=\textwidth]{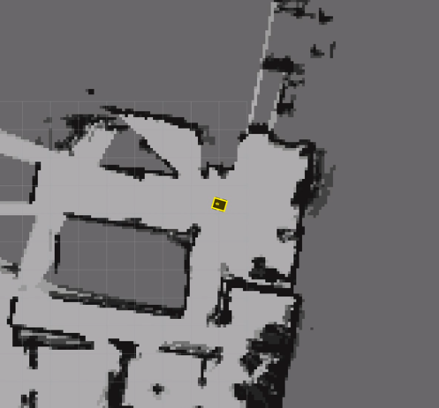}
        \caption{\label{fig:endmap}LIDAR map at the end of a trial. Dark gray are areas unscanned by the LIDAR.}
    \end{subfigure}
    \begin{subfigure}[t]{0.24\textwidth}
    \centering
        \includegraphics[width=0.8\textwidth]{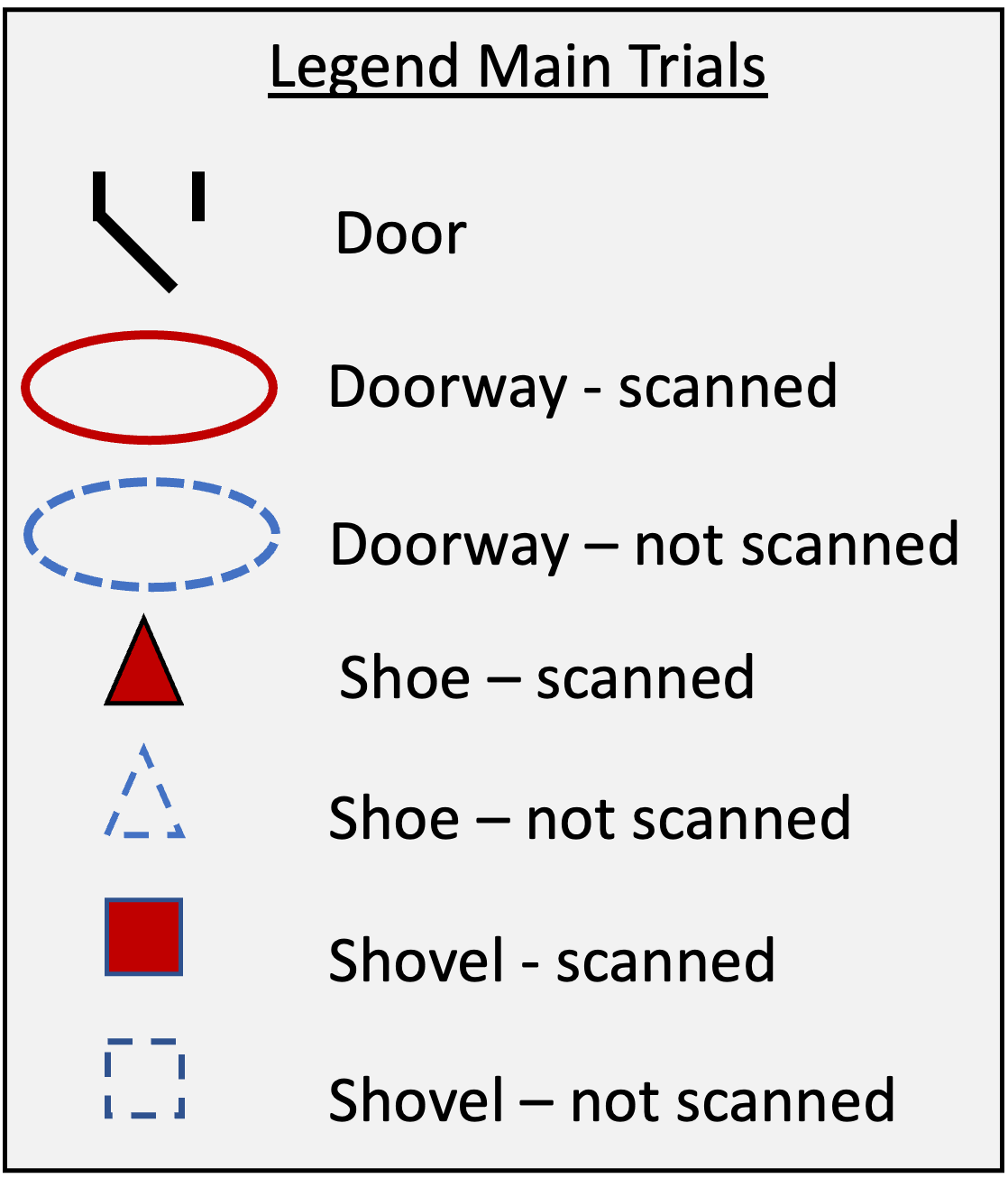}
        \caption{\label{fig:legend}Legend}
    \end{subfigure}
    \begin{subfigure}[t]{0.37\textwidth}
        \includegraphics[width=\textwidth]{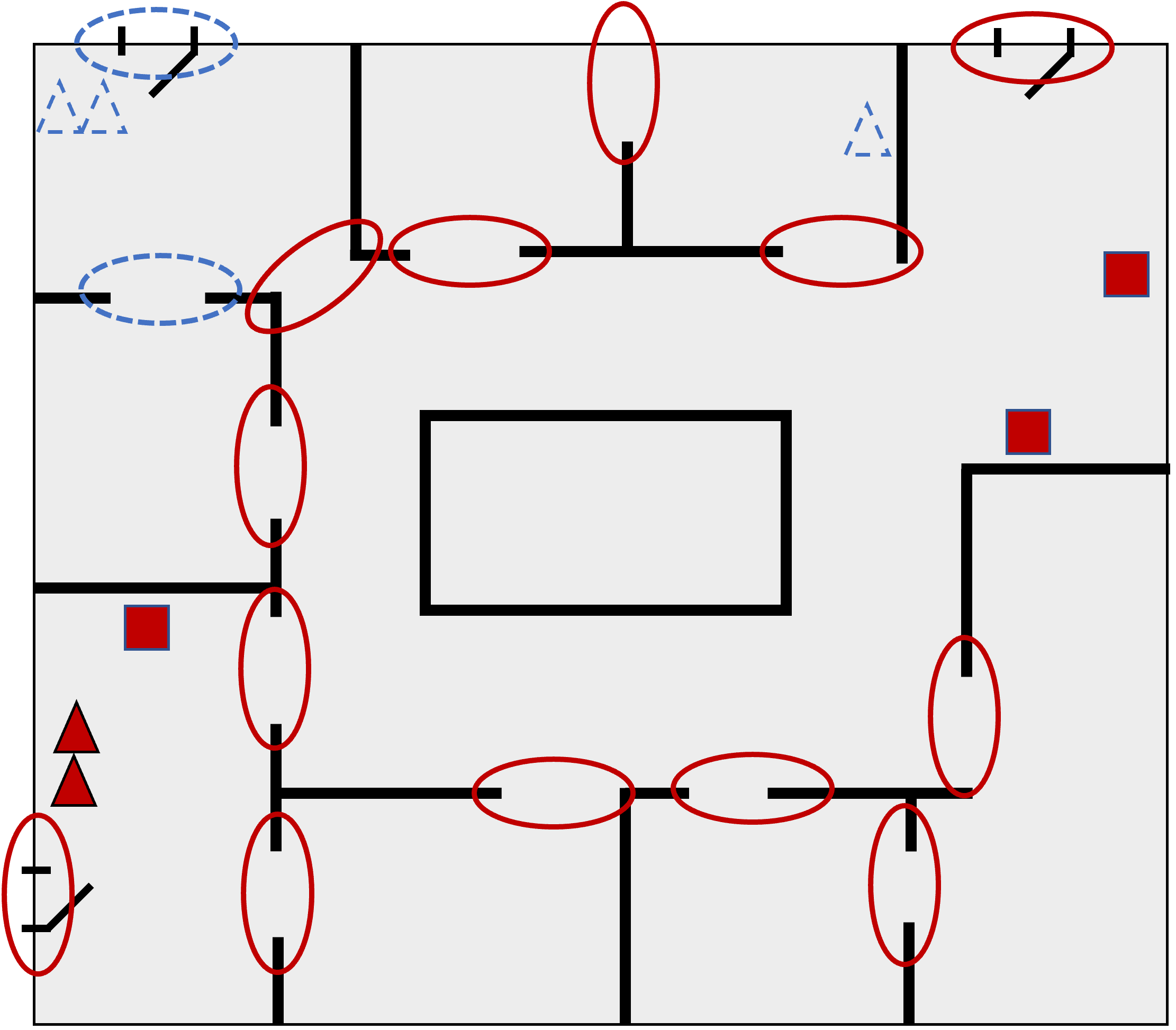}
        \caption{\label{fig:floorplan}Corresponding top-down floor plan of Fig~\ref{fig:endmap} annotated with which items were scanned by the LIDAR.}
    \end{subfigure}
    \caption{One LIDAR map and annotated floor plan with items scanned or not scanned by the LIDAR marked. Floor plan and legend were not shown during the exercise.}
    \label{end-map}
\end{figure*}

\section{Quantifying Success}
\label{sec:success}

Our treatment of success in this multi-modal exercise is the participant's performance on the doorway and objects-of-interest counting tasks in the two main trials. We compute two types of counting success measures, an absolute success and a relative success. 
The absolute success is the number of doorways and objects of interest the participant tallied divided by the total number of these items in the environment. There were three shovels, five individual shoes, and fifteen doorways. The \textit{Absolute Doorway Success} for a participant who tallied 8 doorways would be $8/15 = 53\%$. 
The relative success, on the other hand, differentiates the ability to explore the whole space from the thoroughness with which explored areas were investigated. It measures how well the participants inspected the area they did explore, rather than considering items that were in regions of the environment that they did not get to. Thus, the relative success is the ratio of items tallied divided by the number of items the LIDAR {\it scanned} in the particular trial. 

Fig.~\ref{fig:endmap} shows the LIDAR map at the end of one participant's trial. For each trial, an annotated floor plan was created that denotes the location of all the doorways, shovels, and shoes, and differentiates between the items which were scanned by the LIDAR and those that were not. The annotated map corresponding with Fig.~\ref{fig:endmap} is shown in Fig.~\ref{fig:floorplan}. Items that were scanned by the LIDAR are marked in solid red lines, and those not scanned are in dashed blue (see Fig.~\ref{fig:legend} for the legend; no such figure was given to the participant.) The determination of scanned or not scanned is based on if the LIDAR shows light gray or dark gray, as dark gray denotes areas not scanned. The participant who created this LIDAR map passed thirteen doorways. If they reported tallying eight doorways, their {\it Relative Doorway Success} would be $8/13 = 62\%$. 

Doorway and object successes are assessed separately as these tasks relied on different modalities: shoe and shovel counts required the use of photos; doorway counts could be achieved via LIDAR and/or photos. In practice, we often observed participants counting doorways directly from their final LIDAR map. 
There is the possibility for a false positives when participants are tallying, whether because they got disoriented on the map and counted the same item twice, or because they misidentified a different item for the target item. Our measure does not ensure mistakes of the latter are not taking place as participants were not instructed to verbally note when they found an item, but to address the former, we deduct a $0.5$ penalty when participant counted more than the absolute or their relative total (5 out of 40 count tasks). 
In addition to the success of counting doorways and objects, we are also interested in a participant's success exploring the space. \textit{Area Explored} is critical in real-world tasks such as search-and-rescue, so we scored on percentage of the space the participant covered, as measured by the number of rooms and hallways the LIDAR fully scanned. For the LIDAR map in Fig.~\ref{fig:endmap}, the {\it Area Explored} is $3/4$ hallways (the leftmost hallway was not fully scanned) and $5/9$ rooms; $8/13 = 62\%$.

\section{Exercise Summary}
\label{sec:summary}

We analyzed ten participants who completed two main trials (female=2, average age=43.9) \cite{marge2016applying}. 
For the doorway counting task, the average \textit{Absolute Doorway Success} was 58\% 
and the average \textit{Relative Doorway Success} was 69\%. 
Different participants scored the highest in \textit{Absolute Doorway Success} and \textit{Relative  Doorway Success}. 
In the object (shovel or shoe) counting task, participants scored on average a 61\% \textit{Absolute Object Success} 
and a 68\% \textit{Relative Object Success} (summarized in Table~\ref{tab:stats} with minimum and maximum range).
Three of the trials that had achieved less than a 100\% \textit{Absolute Object Success}, achieved 100\% in the \textit{Relative Object Success.}
The average \textit{Area Explored} was 67\%.

\begin{table}[!th]
    \centering
    \caption{Average success scores and characteristics across twenty trials (two per participant). Right-most column indicates for which hypotheses the variable is tested.} 
    \label{tab:stats}
    \begin{tabular}{lll}
    \toprule
        ~ & Average (min-max) & Hypotheses \\ \midrule
        Counting Task Success & ~ & H$_1$, H$_2$, H$_5$ \\
        \hspace{0.1in}Absolute Doorway Success & 58\% (33-97\%) \\
        \hspace{0.1in}Relative Doorway Success & 69\% (33-100\%) \\
        \hspace{0.1in}Absolute Object Success & 61\% (43-100\%) \\
        \hspace{0.1in}Relative Object Success & 68\% (33-100\%) \\ \midrule
        Area Explored & 67\% (31-100\%) & H$_1$, H$_4$, H$_6$ \\ \midrule
        Photos Taken & 34.5 (13-88) & H$_2$, H$_4$, H$_7$ \\
        Photo Request Strategy & ~ & H$_3$\\
        \hspace{0.1in}Front & 12 trials \\
        \hspace{0.1in}Cardinal & 2 trials\\        \hspace{0.1in}360\textdegree & 2 trials\\
        \hspace{0.1in}Mixed & 4 trials\\ \midrule
        TUs & 34.4 (18-56) & H$_5$, H$_6$, H$_7$ \\
        \bottomrule
    \end{tabular}
\end{table}

The average number of \textit{Photos Taken} in each trial was 34.5. We additionally summarize different {\it Photo Requesting Strategies} exhibited during the exercises: 
\begin{itemize}
    \item Front: a request for a single, forward-facing photo, typically after the robot completes an activity (e.g., ``move forward five feet then take a photo'')
    \item Cardinal: a request for four photos in the cardinal directions (e.g., ``take pictures in north south east and west directions'')
    \item 360\textdegree: a request to take a sequence of photos in a circle (e.g., ``pivot three hundred and sixty degrees to the right taking a picture every forty five degrees'')
\end{itemize}
Twelve trials consistently used the Front strategy, two the Cardinal, two the 360\textdegree, and four trials mixed two or more of the strategies.
The average number of \textit{TUs} per trial, where each TU signifies a new speaker intent, was 34.4.

\section{Analyses and Discussion}
\label{sec:discussion}

To test for relationships, we used a series of multiple regressions that included participant ID (dummy-coded) to control for individual participant effects. Regressions were performed to examine the effects of \textit{Area Explored} on \textit{Absolute Success} (H$_1$), the effects of number of \textit{Photos Taken} on \textit{Absolute and Relative Success} (H$_2$), the effect of \textit{Photo Requesting Strategies} on \textit{Absolute and Relative Success} (H$_3$), the effects of number of \textit{Photos Taken} on \textit{Area Explored} (H$_4$), the effects of \textit{TUs} on \textit{Absolute and Relative Success} (H$_5$) and \textit{Area Explored} (H$_6$), and the effects of \textit{TUs} on \textit{Photos Taken} (H$_7$). Table~\ref{tab:stats} further denotes which variables are tested within each hypothesis.

\subsection{More exploration is good for finding doorways}

H$_1$ expresses a seemingly straightforward claim: increased exploration of the space leads to better performance on item counting tasks. If supported, this would suggest participants have the potential to be more successful if they prioritize exploring  the space in full, if even cursorily, instead of other exploration strategies. Regression controlling for participant ID (overall model $R^2$ = 0.85, F(10, 9) = 4.90, p $<$ 0.001) found a marginal effect of \textit{Area Explored} on \textit{Absolute Doorway Success} (B = 0.24, p = 0.053), suggesting that exploring more area may improve the ability to find and count doorways. This relationship was not found for \textit{Absolute Object Success}.

The positive prediction between higher \textit{Area Explored} and \textit{Absolute Doorway Success} was expected, because exploring more space should allow a participant to encounter and notice more of the doorways in the building. It is somewhat surprising that this relationship was not also found with \textit{Absolute Object Success.} However, the total number of doorways in the building provided a greater range of potential scores for doorway counts (in contrast to total available objects to count), making effects of \textit{Area Explored} easier to detect. Future work with more objects may elucidate whether \textit{Area Explored} improves object scores.

\subsection{Effects of photos taken and photo strategies on counting success and exploration}

We developed H$_2$ and H$_4$ to explore the relationship between the photo modality and the item-counting success and area exploration respectively, and H$_3$ to explore the relationship between {\it Photo Requesting Strategies} and item-counting success. The intuition behind H$_2$ is that since objects can only be seen from photos, taking more photos has the potential to reveal more objects. For \textit{Absolute} and \textit{Relative Object Success}, this was not supported, nor was it for \textit{Relative Doorway Success}.
However, we did find evidence that amount of \textit{Photos Taken} significantly predicted \textit{Absolute Doorway Success} (B = 0.007, p = 0.048), after controlling for participant effects (overall regression $R^2$ = 0.93, F(10, 9)=12.52, p $<$ .001). 
As in H$_1$, other relationships may not have been found because there were so few objects in the environment, and thus not much room for variability in the scores.

In H$_3$, we expected that {\it Photo Requesting Strategies} that were more systematic and did not require the robot to closely examine every particular item in the room, i.e., 360\textdegree\hspace{0.01in} and Cardinal, might lead to higher potential for seeing objects and doorways, and therefore, yield higher {\it Success Scores}. However, our sample size was too small to statistically decouple participants from {\it Photo Requesting Strategies}, so instead we report general observations. The user of the 360\textdegree\hspace{0.01in} strategy did indeed achieve very high scores for \textit{Area Explored} and \textit{Absolute Object Success}. Interestingly, two participants who both began their first trial with a Front strategy later experimented with a Cardinal strategy. They both might have deemed this to be more effective, as their entire second trial used the Cardinal strategy. 

H$_4$ posits that instead of a positive relationship, there may be a trade-off between requesting photos and covering new ground during a trial, such that prioritizing one over the other may negatively effect success. Requesting photos and viewing them takes time, and this time or effort may inhibit the participant's ability to fully cover the space. We therefore might expect participants who request more photos to cover less of the building space in the allotted time. Regression revealed that the number of \textit{Photos Taken} is significantly positively related to \textit{Area Explored} (B = 0.02, p = 0.010) after controlling for participant (overall regression model $R^2$ = .79, F(10,9) = 3.35, p = 0.042).
We were somewhat surprised to find no evidence that taking photos slowed area exploration. However, the significant, positive relationship suggests that the more areas a person explores, the more need they have for additional photos to be taken to understand the new space. 
This also suggests that taking photos is not a hindrance to area exploration. In our exercise, participants could continue to view the previous photo up until they requested a new one, and this viewing could be done concurrently with additional robot movement. The overall time costs, then, may have been trivial.

\subsection{Effects of dialogue on task success and area exploration}

We developed H$_5$ and H$_6$ to explore the relationship between the natural language modality (i.e., total number of TUs) and the item-counting success and area exploration respectively. We posit that each new TU contributes to the participant's understanding of the environment, either by issuing instructions well enough to gain knowledge to complete the counting tasks, or by issuing instructions that send the robot to navigate further, therefore increasing area coverage. 
Total \textit{TUs} helped predict \textit{Absolute Doorway Success} (B = -0.01, p = 0.019) after controlling for individual participant (overall regression $R^2$ = .94, F (10,9) = 15.3, p $<$ 0.001) but had no significant effect on other item-counting successes nor on \textit{Area Explored}. 

As TUs varied in number of turns, and overall signify the number of instigated intentions, the negative relationship between lower {\it Absolute Doorway Success} and higher {\it TUs} is {\it not} revealing that trials with more TUs had a high number of miscommunications (i.e., more turns spent clarifying, therefore, achieving less doorway counting success); instead, this relationship might suggest that participants who focused more on language communication (i.e., issued more instructions) devoted less focus to the LIDAR map (which was the clearest way to count doorways). The non-significant relationship between TUs and the other measures of success could suggest that using a balance of modalities was a more effective strategy than over-relying on natural language alone. We reflect more on this in the final hypothesis below.

\subsection{Relationship between dialogue and photos}

H$_7$ posits a relationship between the natural language and photograph communication modalities. Do participants who take more photos---where each request for a photo is a new intent---issue more intents overall? Or would preference for one modality reduce the use of the other? Results revealed no significant relationship between \textit{TUs} and \textit{Photos Taken}, thus showing no clear evidence of a trade off. 

In reflecting on the choice of {\it TU} as a measure in H$_7$, as well as in H$_5$ and H$_6$, the number of \textit{TUs} alone may not be a sufficiently nuanced measure for how participants chose to use natural language to interact with the robot, and subtle synergies across modalities may be more important factors.
In the following exchange, for instance, we observe a disconnect between the participant's mention of an ``overhead light'' and the robot's knowledge of the space:

\begin{itemize}
\item[] CMD: ``robot continue down the hallway directly in front of you underneath the overhead light''
\item[] DM: ``I don't see an overhead light in my current position.  Would you like me to send a photo?''
\item[] DM: ``robot send a photo''
\item[] --- (photo is sent)
\item[] CMD: ``robot continue moving forward to the right of the red bucket''
\end{itemize}

Once the photo is sent, however, the participant sees indeed there is no overhead light reference in view---which may have been based on a faulty assumption of what was seen in the prior photo sent which was not the current location of the robot in the LIDAR map---and so issues a new intent surrounding the `red bucket' in view. This exchange is comprised of 2 TUs, yet the number alone does not capture the nuance of what is happening. Thus in future work we will explore these instances through other measures.

\section{Comparisons to Other Collaboration Paradigms}
\label{sec:related}

There are many ways in which a human and a robot can remotely work together. We surveyed a number of human-robot collaborative exercises, and more broadly, human-human dyads that follow the roles of \textit{director} (e.g., our participant) and \textit{follower} (e.g., our robot). From these experimental designs and constraints, we highlight the following dimensions that characterize the exercises: 
\begin{itemize} 
\item the task (navigating through an environment, configuring an environment);
\item the environment (real world, simulated world, static image);
\item the modality information provided; 
\item the timing and extent to which information is made available and shared with the director, including if they are given a complete environment or map to start with, or whether this information must be discovered from information provided by the follower.
\end{itemize}
Our exercise is similar in many ways to previous tasks reported in the literature, but it addresses a realistic and under-explored part of this feature space, for which our analyses provide a baseline understanding.

In many prior works, the director in the dyad was provided with a complete map (static or dynamic) of the environment that the follower occupied \cite{liu2016coordinating,de2018talk,eberhard2010indiana,padmakumar2021teach,stoia2008scare,suhr2019executing,gervits2021should}. With their knowledge of the environment layout, and in some cases, its contents, the director could give focused instructions to orient and steer the follower to the correct destination. In contrast, our participants directed the robot through the environment {\it in order to} explore and so come to understand the space by expanding the LIDAR map and taking pictures. Because the participants did not have the complete map when they began, the {\it Area Explored} and {\it Relative Success} scores provide a new way to capture just how much area is explored during navigation tasks 
and the effect of that exploration on task success. 

Like other real-world collaborative tasks (e.g, \cite{liu2016coordinating,de2018talk,eberhard2010indiana}), our participant did not receive visual information from the robot's perspective via streaming, which is commonly afforded to directors whose followers occupy a simulated environment (e.g., \cite{padmakumar2021teach,stoia2008scare,suhr2019executing,gervits2021should}). In our paradigm, our participants had neither the real-time streaming perspective of the visual, nor a complete lack of visual information. By allowing our participants to request photos on demand, our {\it Photo Requesting Strategies} provide insights into the different ways that participants interacted with the photo modality. Such interactions cannot be as readily measured through continuous streaming where the participant does not have to initiate a request to view the modality. As revealed through the dialogue excerpt in the `Relationship between dialogue and photos' discussion above, the lack of common ground is apparent when the photo and LIDAR modalities are not in sync, a phenomena that is again a result of the participant having neither real-time streaming, nor a complete lack of visual information. Furthermore, our analysis of the positive relationship between the number of photos taken and {\it Area Explored} suggests that participants can still explore and gain insights into the environment without having the continual visual stream. 

\section{Conclusions and Future Work} \label{sec:future}

The present work investigates the relationship between task success and the use of three communication modalities--language, LIDAR, and photos--in a human-WoZ-robot exercise. 
We found evidence that increased use of photos may have helped participants identify and count doorways in the remote environment. We also found evidence that requesting photos does not appear to hinder the amount of overall area exploration. This is encouraging, as it suggests that robot Commanders do not need to sacrifice the amount of remote area covered in favor of requesting more robot-POV images of that area. Different photo-taking strategies may also be associated with different levels of task success. For example, the person who used a 360\textdegree\hspace{0.01in}  photo strategy was able score well on object-of-interest counts $and$ total area exploration, suggesting that taking many photos, in a language-efficient way, may provide benefits both for \textit{Area Explored} and focused object identification within that area. Future analyses on more participants may reveal additional relationships and further inform default robot strategies or other design parameters.  

To enable further exploration of multi-modal communication strategies and design in human-robot navigation, we release our unique multi-modal human-robot communication corpus, \textit{SCOUT}. The corpus includes language used across all dialogue floors along with the associated still photos taken and LIDAR maps. We plan to make this available at \url{github.com/USArmyResearchLab} or by contacting the first author.

In future work, we look towards automating the DM-Wizard so it can react based on data-driven policies, especially with respect to the language modality when miscommunication occurs within a TU, or a recommendation for navigation strategies to increase the space seen. We will investigate how task success and modality characteristics change across varied experimental designs, including the use of photo strategies and space seen in a virtual facsimile of the building. Additionally, we plan to design new ways for interacting with this multi-modal data. We currently have screen recordings of trials from the participant display (Fig.~\ref{exp-view}) to replay interactions; however, of greater utility would be a complete transcript artifact incorporating photo and LIDAR modalities into the dialogue language by conversational floor in Table~\ref{tab:tu}. We are working to create this artifact with the expectation that it will benefit the broader multi-modal communication research community.

\section{Limitations}

Due to our choice of experimental design, our collected data may not generalize. For instance, the low-lighting in the images may prove challenging for SotA computer vision  algorithms trained on `canonical' environments. 



\bibliographystyle{IEEEtran}
\bibliography{root}


\end{document}